\documentclass{elsart}
\usepackage[dvips]{graphicx}
\newcommand{\kmunudk}{\ensuremath{D^0 \rightarrow K^- \mu^+ \nu }}
\newcommand{\krz}{\ensuremath{K^{*0}}}
\newcommand{\krzb}{\ensuremath{\overline{K}^{*0}}}
\newcommand{\krzmndk}{\ensuremath{D^+ \rightarrow \krzb \mu^+ \nu}}

\newcommand{\philndk}{\ensuremath{D_s^+ \rightarrow \phi\; \ell^+ \nu_\ell}}

\newcommand{\kpimndk}{\ensuremath{D^+ \rightarrow K^- \pi^+ \mu^+ \nu }}

\newcommand{\gevcsq}{\ensuremath{\textrm{GeV}/c^2}}
\newcommand{\gevcqt}{\ensuremath{\textrm{GeV}^2/c^4}}

\newcommand{\thv}{\ensuremath{\theta_\textrm{v}}}
\newcommand{\thl}{\ensuremath{\theta_\ell}}
\newcommand{\costhv}{\ensuremath{\cos\thv}}
\newcommand{\costhvsq}{\ensuremath{\cos^2\thv}}
\newcommand{\sinthv}{\ensuremath{\sin\thv}}
\newcommand{\costhl}{\ensuremath{\cos\thl}}

\newcommand{\sinthl}{\ensuremath{\sin\thl}}
\newcommand{\sinthlsq}{\ensuremath{\sin^2\thl}}
\newcommand{\qsq}{\ensuremath{q^2}}

\newcommand{\bw}{\ensuremath{\textrm{BW}}}
\newcommand{\mkpi}{\ensuremath{m_{K\pi}}}

\newcommand{\rtwo}{\ensuremath{r_2}}
\newcommand{\rthree}{\ensuremath{r_3}}
\newcommand{\rvee}{\ensuremath{r_v}}
\newcommand{\mysection}[1]{\section{#1}}
\newcounter{saveeqn}%

\begin{document}
\begin{frontmatter}
\title{A Non-parametric Approach to the $\krzmndk$ Form Factors}
\collaboration{The FOCUS Collaboration}

\author[ucd]{J.~M.~Link}
\author[ucd]{P.~M.~Yager}
\author[cbpf]{J.~C.~Anjos}
\author[cbpf]{I.~Bediaga}
\author[cbpf]{C.~G\"obel}
\author[cbpf]{A.~A.~Machado}
\author[cbpf]{J.~Magnin}
\author[cbpf]{A.~Massafferri}
\author[cbpf]{J.~M.~de~Miranda}
\author[cbpf]{I.~M.~Pepe}
% ...

\address[ucd]{University of California, Davis, CA 95616}
\address[cbpf]{Centro Brasileiro de Pesquisas F\'{\i}sicas, Rio de Janeiro, RJ, Brasil}
% ...
\nobreak
\begin{abstract}
Using a large sample of $D^+ \rightarrow K^- \pi^+ \mu^+ \nu$ decays
collected by the FOCUS photoproduction experiment at Fermilab, we
present the first measurements of the helicity basis form factors free
from the assumption of spectroscopic pole dominance. We also present the
first information
on the form factor that controls the $s$-wave interference discussed in
a previous paper by the FOCUS collaboration. We find reasonable agreement with the
usual assumption of spectroscopic pole dominance and measured form factor
ratios.

\end{abstract}
\end{frontmatter}

% No page number printed for this page
%\tableofcontents    % contents are based on the sections, subsections, etc.
%\listoffigures     % based on the figures

\mysection{Introduction}

The \kpimndk{} decay is described in terms of helicity basis form
factors that give the \qsq{} dependent amplitudes for the \krzb{} to
be in any of its possible angular momentum states~\cite{KS}.
Traditionally~\cite{formfactor,otherff}, these helicity basis form factors
are written as linear combinations of vector and axial form factors
that, in turn, are assumed to have a \qsq{} dependence given by
spectroscopic pole dominance. The pole masses are fixed to the 
known masses of the excited $D_s^+$ states with vector and axial quantum
numbers. This paper uses a new weighting technique to disentangle and
directly measure the \qsq{} dependence of these helicity basis form
factors free from the assumption of spectroscopic pole dominance. We
believe this paper represents the first non-parametric analysis of the
\kpimndk{} helicity basis form factors.

\begin{figure}[tbph!]
 \begin{center}
  \includegraphics[width=3.0in]{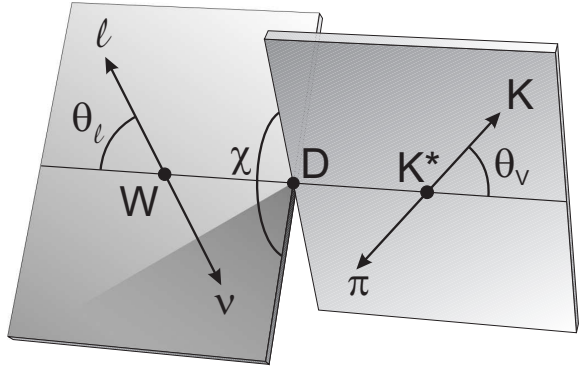}
  \caption{Definition of kinematic variables.
 \label{angles}}
 \end{center}
\end{figure}

Five kinematic variables that uniquely describe $\kpimndk$ decay are
illustrated in Figure~\ref{angles}. These are the $K^- \pi^+$
invariant mass ($\mkpi$), the square of the $\mu\nu$ mass ($\qsq$),
and three decay angles: the angle between the $\pi$ and the $D$
direction in the $K^- \pi^+$ rest frame ($\thv$), the angle between
the $\nu$ and the $D$ direction in the $\mu\nu$ rest frame ($\thl$),
and the acoplanarity angle between the two decay planes
($\chi$). 
The full intensity distribution for $\kpimndk$, differential in 
these five kinematic variables, is given in Reference~\cite{formfactor}. 
Throughout this paper, we will use the simpler expression given by Eq.~\ref{amp1} that  
gives the form of this decay intensity after integration over the acoplanarity angle $\chi$.
The $\chi$ integration significantly simplifies the intensity by eliminating all interference
terms between different helicity states of the virtual $W^+$ with relatively little loss in form factor
information.\footnote{The acoplanarity acceptance in our spectrometer is uniform to within 2\%, hence
no correction is required given the size of our statistical errors.}
Following the 
notation of~\cite{KS}, the Eq.~\ref{amp1} intensity is written in terms of
the four helicity basis form factors:
$H_+(\qsq), H_0(\qsq), H_-(\qsq), H_t(\qsq)$ along with an additional form factor $h(\qsq)$ that describes the coupling
of an additional, small $s$-wave amplitude contribution (with a constant
amplitude $Ae^{i \delta}$) that is discussed in Reference~\cite{anomaly}. 
Our earlier papers~\cite{anomaly,formfactor} assumed that the s-wave amplitude had the same
zero helicity form factor as that for $\krzmndk$ -- {\emph e.g.} $h_0(\qsq) = H_0(\qsq)$.\footnote{Eq.~\ref{amp1}
drops the term which is second order in the small amplitude $A$.} The acoplanarity averaged intensity in terms of 
the decay angles, helicity form factor products, and \krzb{} Breit-Wigner amplitude ($BW$) is:  
\begin{eqnarray}
   \int { |A|^2 d\chi }
      = \frac{q^2 - m_\ell ^2}{8} \left\{ \begin{array}{l}
              ((1 + \costhl) \sinthv)^2 |H_+(q^2)|^2 |\bw|^2  \\ 
            + ((1 - \costhl) \sinthv)^2 |H_-(q^2)|^2 |\bw|^2  \\ 
            + (2 \sinthl \costhv)^2 |H_0(q^2)|^2 |\bw|^2  \\ 
            + 8 \sinthlsq \costhv H_0(q^2)h_o(q^2)
                {\mathop{\rm Re}\nolimits}\{Ae^{-i\delta} \bw \}
            \end{array} \right\} \nonumber \\
  \mbox{} \label{amp1} \\
  \mbox{} +\frac{|\bw|^2}{8}(q^2 - m_\ell^2)\frac{m_\ell^2}{q^2}
                                 \left\{ \begin{array}{l}
              (\sinthl \sinthv)^2 |H_+|^2 
                + (\sinthl \sinthv)^2 |H_-|^2  \\ 
            + (2 \costhl \costhv)^2 |H_0|^2  \\ 
            + (2 \costhv)^2 |H_t|^2
            + 8\costhl \costhvsq H_0 H_t
            \end{array} \right\} \,, \nonumber
\end{eqnarray}
where 
\begin{equation}
\bw = \frac{\sqrt{m_0 \Gamma} \left(\frac{P^*}{P_0^*}\right)}
	   {m_{K\pi}^2 - m_0^2 + i m_0 \Gamma 
	      \left(\frac{P^*}{P_0^*}\right)^3} \,.
\end{equation}

The first term gives the intensity for the $\mu^+$ to be
right-handed, while the (highly suppressed) second term gives the
intensity for it to be left-handed.\footnote{We are using a $p$-wave
Breit-Wigner form with a
width proportional to the cube of the kaon momentum in the kaon-pion
rest frame ($P^*$) over the value of this momentum when the kaon-pion
mass equals the resonant mass ($P^*_0$).  The squared modulus of our
Breit-Wigner form will have an effective $P^{*3}$ dependence in the
numerator as well. Two powers $P^*$ come explicitly from the $P^*$ in
the numerator of the amplitude and one power arises from the 4 body
phase space.}

Before describing the non-parametric approach,
we begin with a description of the traditional experimental approach.
The $\krzmndk$ decay amplitude is typically 
analyzed~\cite{KS} in terms of four form factors. This intensity
expression is written in terms of four helicity basis form factors that are in turn written as linear combinations of vector and axial form factors as given in Eq.~\ref{helicity}.
\begin{eqnarray}
H_\pm(\qsq) &=&
   (M_D+\mkpi)A_1(\qsq)\mp 2{M_D K\over M_D+m_{K\pi}}V(\qsq) \,,
                               \nonumber \\
H_0(\qsq) &=&
   {1\over 2\mkpi\sqrt{\qsq}}
   \left[
    (M^2_D -m^2_{K\pi}-\qsq)(M_D+\mkpi)A_1(\qsq) \frac{}{} \right. \nonumber \\
          & & \mbox{} \hspace{2cm} \left.
    -4{M^2_D K^2\over M_D+\mkpi}A_2(\qsq) \right] \,,\label{helicity} \\ 
H_t(\qsq) &=&
   {M_D K\over m_{K\pi}\sqrt{\qsq}}
   \left[ (M_D+m_{K\pi})A_1(\qsq)
    -{(M^2_D -m^2_{K\pi}+\qsq) \over M_D+m_{K\pi}}A_2(\qsq) \right. \nonumber\\
          & & \mbox{} \hspace{2cm} \left.
    +{2\qsq\over M_D+m_{K\pi}}A_3(\qsq) \right] \,,\nonumber 
\end{eqnarray}
where $K$ is the momentum of the $K^- \pi^+$ system in the rest frame
of the $D^+$.
The vector and axial form factors are generally parameterized by a pole
dominance form:
\begin{equation}
A_i(\qsq)={A_i(0)\over 1-\qsq/M_A^2}~~~~\textrm{and}~~~~
V(\qsq)={V(0)\over 1-\qsq/M_V^2} \,,
\end{equation}
where all previous experiments used {\it spectroscopic} pole masses of $M_A =
2.5~\gevcsq$ and $M_V = 2.1~\gevcsq$ which in the case
of $\krzmndk$ are tied to masses of vector and axial $D_s^*$ states.

Using the spectroscopic pole dominance assumption, previous experiments~\cite{formfactor,otherff} have fit the shape of the $\krzmndk$ intensity 
to at most 3 parameters which are ratios of form factors taken at $\qsq = 0$\,: $\rvee \equiv V(0)/ A_1(0),\ \rtwo \equiv A_2(0)/A_1(0),\ \rthree \equiv A_3(0)/A_1(0)$ and in one case~\cite{formfactor} the (constant) $s$-wave complex amplitude. 

As in our earlier paper~\cite{qsq} on the $\qsq$ dependence of the $\kmunudk$ form factor, 
we present the first non-parametric measurements of the helicity basis form
factors that 
describe $\kpimndk$ decay.  In particular, we will provide information on 
$H_\pm^2(\qsq)$, $H_0^2(\qsq)$, and $h_0(\qsq) \times H_0(\qsq)$ in bins
of $\qsq$ by projecting out the associated angular factors given in 
Eq.~\ref{amp1}. The cross term $h_0(\qsq) \times H_0(\qsq)$ represents
the interference between the $s$-wave and the $\krz$.

Throughout this paper, unless explicitly stated otherwise,
the charge conjugate is also implied when a decay mode of a specific
charge is stated.

\mysection{Experimental and analysis details}

The data for this paper were collected in the Wideband photoproduction
experiment FOCUS during the Fermilab 1996--1997 fixed-target run. In
FOCUS, a forward multi-particle spectrometer is used to measure the
interactions of high energy photons on a segmented BeO target. The
FOCUS detector is a large aperture, fixed-target spectrometer with
excellent vertexing and particle identification. The FOCUS
experiment and analysis techniques have been described
previously~\cite{anomaly,nim,CNIM}.

To isolate the
$\kpimndk$ topology, we required that candidate muon, pion, and kaon
tracks formed a secondary vertex with a confidence level
exceeding 25\%. We required a primary vertex consisting of 
at least two charged tracks. The muon track, when extrapolated to the shielded muon
arrays, was required to match muon hits with a confidence level
exceeding 5\%. The kaon was required to have a \v Cerenkov light
pattern more consistent with that for a kaon than that for a pion by 1
unit of log likelihood~\cite{CNIM}. No \v Cerenkov requirement was made
on the pion. 

To further reduce muon misidentification, a muon candidate was allowed
to have at most one missing hit in the 6 planes comprising our inner
muon system and a momentum exceeding 10 GeV$/c$.  In order to suppress
muons from pions and kaons decaying within our apparatus, we required
that each muon candidate had a confidence level exceeding 2\% to the
hypothesis that it had a consistent trajectory through our two
analysis magnets.

Non-charm and random combinatoric
backgrounds were reduced by requiring both a detachment between the
vertex containing the $K^-\pi^+\mu^+$ and the primary production
vertex of at least 10 standard deviations and a minimum visible energy
$(E_K+E_\pi+E_\mu)$ of 30 GeV. To suppress possible backgrounds from
higher multiplicity charm decays, we isolate the $K\pi\mu$ vertex from
other tracks in the event (not including tracks in the primary vertex)
by requiring that the maximum confidence level for another track to
form a vertex with the candidate be less than 0.1\%.

\begin{figure}[tbph!]
 \begin{center}
  \includegraphics[width=3.in]{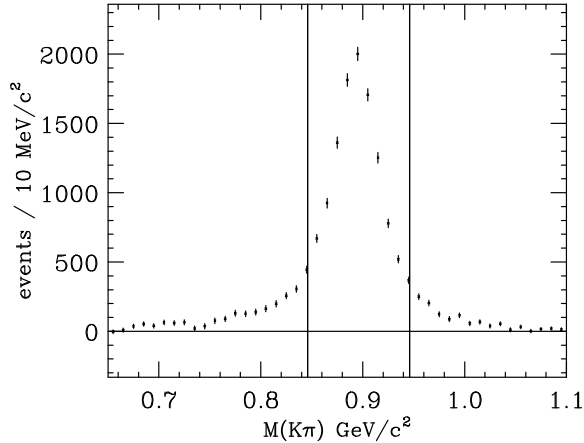}
  \caption{Wrong-sign subtracted $\kpimndk$ signal. Over the full displayed mass range there is a right-sign excess of
  14\,798 events. For this analysis, we use a restricted mass range from 0.846--0.946 $\gevcsq$ (shown by vertical lines).
 In this restricted region, there is a right-sign excess of 11\,397 events.  
\label{signal}} \end{center}
\end{figure}

In order to allow for the missing energy of the neutrino in this
semileptonic $D^+$ decay, we required the reconstructed $K \pi \mu$
mass be less than the nominal $D^+$ mass.  Background from $D^+
\rightarrow K^- \pi^+ \pi^+$, where a pion is misidentified as a muon,
was reduced using a mass cut: we required that when the muon track is
treated as a pion and the combination is reconstructed as a $K \pi
\pi$, the $K \pi \pi$ invariant mass was less than 1.820~$\gevcsq$.
In order to suppress
background from $D^{*+} \rightarrow D^0 \pi^+ \rightarrow (K^- \mu^+
\nu) \pi^+$, we required $M(K^- \mu^+ \nu \pi^+) - M(K^- \mu^+ \nu) >
0.18~\gevcsq $. The wrong-sign subtracted $\mkpi$ distribution for these $\kpimndk$ candidates
is shown in Figure~\ref{signal}. Wrong-sign events have tracks identified 
as $K^- \pi^+ \mu^-$ isolated in a detached vertex.

We will use a restricted mass range from 0.846--0.946 $\gevcsq$ in this analysis in an effort to diminish
the dependence of the helicity basis form factors on $\mkpi$ through Eq.~\ref{helicity}.

The technique used to reconstruct the neutrino momentum through the
$D^+$ line-of-flight and tests of our ability to simulate the
resolution on kinematic variables that rely on the neutrino momentum are
described in Reference~\cite{anomaly}.

\mysection{Projection Weighting Technique} 

In this section, we describe the weighting technique that we use to extract the helicity basis form factors that describe the $\kpimndk$ angular distribution according to Eq.~\ref{amp1}.  
For a given $\qsq$ bin, a weight will be assigned to the event depending on its $\thv$ and $\thl$
decay angle.  We consider 25 
joint  $\Delta \costhv  \times \Delta \costhl$ angular bins:  
5 evenly spaced bins in $\costhv$ times 5 bins in $\costhl$. 
Each event will acquire a weight designed to project out a given helicity form factor that depends on which of the 25 angular bins that it is reconstructed in.
It is convenient to think of weighting as constructing a dot product of the form $\vec P_\alpha \cdot \vec D$ where $\vec D$ is a data vector that consists of the number of events reconstructed in each of the 25 angular vectors 
$\vec D = ( {\begin{array}{*{20}c}
   {n_1} & {n_2} & {...} & {n_{25}}  \\ \end{array}} )$
  and  $\vec P_\alpha  $ is a projection vector for the $\alpha$ helicity form factor. The 25 components of each $\vec P_\alpha$  vector give the weights applied to each event reconstructed in one of the 25 angular bins. Eq.~\ref{likeweighting} says
the product $\vec P_ + \cdot \vec D$ is
equivalent to weighting the events in angular bin 1 by $\left[\vec P_+\right]_1$, 
weighting the events in angular bin 2 by $\left[\vec P_+\right]_2$, etc.

\begin{equation}
\vec P_+ \cdot \vec D = 
    \left[\vec P_+\right]_1 n_1 + \left[\vec P_+\right]_2 n_2 
       + \cdots \left[\vec P_+\right]_{25} n_{25} \,. 
\label{likeweighting}
\end{equation}

The $\vec P_\alpha$ weights are designed to project out the helicity form
factors using Monte Carlo inputs. Here is how they are obtained. To simplify our discussion, consider
the case of just three form factors $H_+^2(\qsq)$,\ $H_-^2(\qsq)$,\
$H_0^2(\qsq)$.
For each $\qsq_i$ bin, let $\{\vec m_\alpha\} = ( {\begin{array}{*{20}c}
{\vec m_+} & {\vec m_-} & {\vec m_0} \end{array}} )$
where $\vec m_\alpha $ is 
the number of events present in each of the 25 angular bins 
when a nominal $H_\alpha(q_i^2)$
is turned on and all other nominal $H_{\alpha \ne \beta}(q_i^2)$ are turned off.  
As indicated in Eq.~\ref{series}, for each
$\qsq_i$ bin the $\vec D_i$ vector can be written as a linear combination of the three $\vec m$ vectors 
with coefficients $f_\alpha(q_i^2)$.
The $f_\alpha(q_i^2)$ functions are proportional to the true $H^2_\alpha(q_i^2)$ 
along with pre-factors such as
$q^2 - m_\ell^2$ and corrections such as acceptance and resolution.  

\begin{equation}
  \vec D_i = f_+(q_i^2)\:\vec m_+ + f_-(q_i^2)\:\vec m_- 
            + f_0(q_i^2)\:\vec m_0 \,.
\label{series}
\end{equation}
We can convert Eq.~\ref{series} to the ``component equation" shown in  
Eq.~\ref{system}.

\begin{equation}
\left( {\begin{array}{*{20}c}
   {\vec m_+ \cdot \vec D_i }  \\
   {\vec m_- \cdot \vec D_i }  \\
   {\vec m_0 \cdot \vec D_i }  \\
\end{array}} \right) = \left( {\begin{array}{*{20}c}
   {\vec m_+ \cdot \vec m_+} & {\vec m_+ \cdot \vec m_-} 
      & {\vec m_+ \cdot \vec m_0 }  \\
   {\vec m_- \cdot \vec m_+} & {\vec m_- \cdot \vec m_-} 
      & {\vec m_- \cdot \vec m_0 }  \\
   {\vec m_0 \cdot \vec m_+} & {\vec m_0 \cdot \vec m_-} 
      & {\vec m_0 \cdot \vec m_0 }  \\
\end{array}} \right) \left( {\begin{array}{*{20}c}
   {f_+(q_i^2)}  \\
   {f_-(q_i^2)}  \\
   {f_0(q_i^2)}  \\
\end{array}} \right) \,.
\label{system}
\end{equation}
The solution to Eq.~\ref{system} can be written as:
\begin{equation}
f_+ (q_i^2) = {}^i\vec P_+ \cdot \vec D_i \,,\
f_- (q_i^2) = {}^i\vec P_- \cdot \vec D_i \,,\
f_0 (q_i^2) = {}^i\vec P_0 \cdot \vec D_i \,,
\label{solution}
\end{equation}
where our notation is that the left superscript gives the \qsq{} bin number
and the subscript identifies the helicity of the projector. 

The ${}^i\vec P_\alpha$ weights are given by Eq.~\ref{projectors}.

\begin{equation}
\left( {\begin{array}{*{20}c}
   {{}^i\vec P_+}  \\
   {{}^i\vec P_-}  \\
   {{}^i\vec P_0}  \\
\end{array}} \right) = \left( {\begin{array}{*{20}c}
   {\vec m_+ \cdot \vec m_+} & {\vec m_+ \cdot \vec m_-}
       & {\vec m_+ \cdot \vec m_0}  \\
   {\vec m_- \cdot \vec m_+} & {\vec m_- \cdot \vec m_-}
       & {\vec m_- \cdot \vec m_0}  \\
   {\vec m_0 \cdot \vec m_+} & {\vec m_0 \cdot \vec m_-}
       & {\vec m_0 \cdot \vec m_0}  \\
\end{array}} \right)^{- 1} \left( {\begin{array}{*{20}c}
   {\vec m_+}  \\
   {\vec m_-}  \\
   {\vec m_0}  \\
\end{array}} \right) \,.
\label{projectors}
\end{equation}

One can correct for acceptance by using the proportionality relations such as
Eq.~\ref{secondweight},

\begin{eqnarray} 
 & f_i^{+} = {}^i\vec P_+ \cdot \vec D_i
         = \frac{ ( H_+(q_i^2) )^2}
                { ( \tilde{H}_i^+)^2}
           \:{}^i\vec P_+ \cdot \vec M_i \nonumber \\
 & \Rightarrow ( H_+(q_i^2) )^2 
         = \left[ \frac{( {}^i\tilde{H}_+ )^2 }
                       { {}^i\vec P_+ \cdot \vec M_i}\:
                  \:{}^i\vec P_+ \right] \cdot \vec D_i
         \equiv {}^i\rho_+ \cdot \vec D_i \,,
\label{secondweight}
\end{eqnarray}

where $\vec M_i$ are the bin populations from a Monte Carlo, generated assuming 
a trial form factor set $\tilde H_+^2(\qsq)$, $\tilde H_-^2(\qsq)$, and $\tilde H_0^2(\qsq)$.
As indicated in Eq.~\ref{secondweight}, the projection weights ${}^i\vec P_+$ and the 
projection-weighted Monte Carlo distributions  (${}^i\vec P_+ \cdot \vec M_i $)
can then be used to construct an adjusted weight vector ${{}^i\vec \rho}_+$.  The (arbitrarily normalized)
form factors $H_+^2(\qsq)$, $H_-^2(\qsq)$ and $H_0^2(\qsq)$ would then be obtained by making three weighted histograms using the ${{}^i\vec \rho}_+$, 
${{}^i\vec \rho}_-$ and ${{}^i\vec \rho}_0$ weights respectively.

We next discuss some of the complications in applying the projective weighting scheme in our experiment due to smearing of kinematic variables because of the missing neutrino.
In the absence of substantial $\qsq$ smearing, a totally arbitrary set of trial form factors 
(such as $\tilde H_\alpha(\qsq_i) = 1 $) can be used to get unbiased estimates of the helicity basis form factors. This is no longer true when smearing 
in the kinematic variables $\qsq,\ \costhv,\ \costhl$ is substantial since one
must use reconstructed quantities in dealing with the data. The acceptance and 
resolution that affect the ${{}^i\vec \rho}_+$, 
${{}^i\vec \rho}_-$ and ${{}^i\vec \rho}_0$ weights
depend on the true kinematic variables. Since the mapping between the true kinematic
variables and reconstructed kinematic variables depends on the underlying form factors, 
one can bias the returned form factors to the extent that  $\tilde H_\alpha(\qsq_i) \ne H_\alpha(\qsq_i)$.
We found through Monte Carlo simulation that it is only important
to get a reasonably good first guess for  $\tilde H_\alpha (\qsq_i)$ and multiple iteration is not required given the size of our statistical error bars. 

Although the mass terms of Eq.~\ref{amp1} are suppressed by two powers of the muon mass, they are surprisingly important at low \qsq{}.
Our analysis includes projectors for each of the six form factor products present in Eq.~\ref{amp1}. Although
we were unable to obtain useful information on $H^2_t(\qsq)$ or the $H_0(\qsq) \times H_t(\qsq)$ interference
term, we needed to allow for them in the construction of ${}^i\vec P_\alpha$ using Eq.~\ref{projectors} to insure that the
projectors used for  $H_\pm^2(\qsq)$, $H_0^2(\qsq)$, and $h_0(\qsq) \times H_0(\qsq)$ will be ``orthogonal"
to the angular terms associated with the $H^2_t(\qsq)$ and $H_0(\qsq) \times H_t(\qsq)$ contributions of Eq.~\ref{amp1}.
Without the incorporation of the mass term projectors, we see a dramatic mismatch between the input and output
form factors for $H_\pm^2(\qsq)$ in the first \qsq{} bin in our Monte Carlo studies. For example, $H_-^2(\qsq)$ nearly drops to zero in the first \qsq{}
bin if projectors for the mass terms are not included.

\begin{figure}[htp]
\begin{center}
\includegraphics[height=3.in]{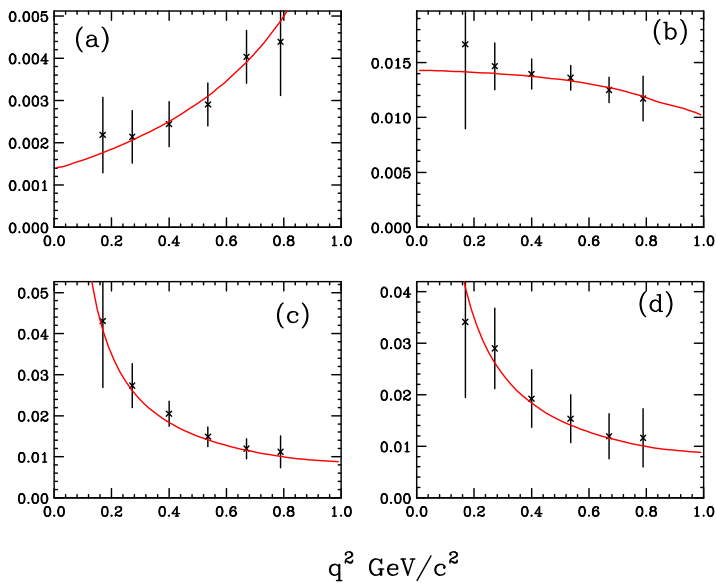}
\end{center}
\caption{ Monte Carlo study of the projective weighting technique. The form factors used in this Monte Carlo simulation are shown as the solid curves.
The reconstructed form factors are the points with error bars. They are plotted in ``arbitrary" units but
the same unit is used for all four form factors in order to convey the relative size of each contribution. The plots are: 
(a)~$H_+^2(\qsq)$,
(b)~$H_-^2(\qsq)$,
(c)~$H_0^2(\qsq)$, and
(d)~$h_0(\qsq) \times H_0(\qsq)$.
\label{ucpop}}
\end{figure}

Figure~\ref{ucpop} summarizes a complete Monte Carlo simulation of the projective weighting technique. This Monte Carlo
was run with 9 times our data sample but we have inflated the error bars by
a factor of three to indicate the estimate of the errors expected in the data.
The form factor measurements are plotted
at the abscissa of the average \emph{generated} $\qsq$ for each of the 6 evenly spaced \emph{measured} $\qsq$ bins
rather than at the measured bin center.  The good agreement between the input and output form factors validates our assumption that the 
$|A|^2$ term can be dropped when constructing the projective weights since these terms are included in the Monte Carlo simulation.

The method does an excellent job at reproducing the input form factor products (shown as a curve) 
both in terms of the shape and relative contribution from each of the 4 form factor products.  The input form factors used the measurements and $s$-wave model developed in  Reference~\cite{formfactor}.  

\begin{figure}[htp]
\begin{center}
\includegraphics[height=3.in]{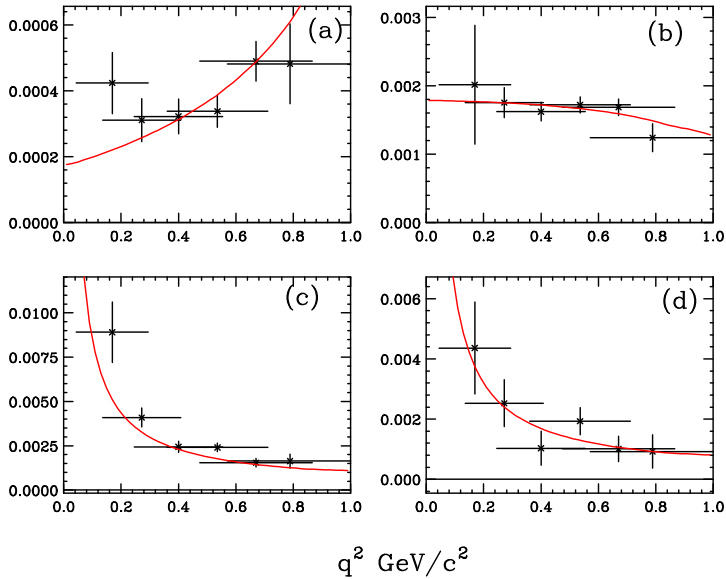}
\end{center}
\caption{
Non-parametric form factors obtained for data with horizontal error bars given by our r.m.s.
resolution. The plots are:
(a)~$H_+^2(\qsq)$,
(b)~$H_-^2(\qsq)$,
(c)~$H_0^2(\qsq)$, and
(d)~$h_0(\qsq) \times H_0(\qsq)$.
The r.m.s $\qsq$ resolution varies (monotonically) from 0.125 $\gevcqt$, in the first bin, 
to 0.218 $\gevcqt$ in the last bin.
\label{ucdataX}}
\end{figure}

Figure~\ref{ucdataX} and Table~\ref{unconvsum} show the results obtained for data.  
The form factor measurements are plotted
at the abscissa of the average \emph{generated} $\qsq$ for each of the 6 evenly spaced \emph{measured} $\qsq$ bins
as determined from the Monte Carlo. The horizontal error bars given the r.m.s \qsq{} resolution for
each bin.
We have subtracted the weighted distribution 
obtained for background events
simulated in our charm Monte Carlo that incorporates all known charm decays.  
Non-charm backgrounds are primarily 
eliminated through a wrong-sign subtraction. 
Applying more stringent cuts in the analysis indicates that the only 
significant deviation from the model occurs in plot a) of Fig.~\ref{ucdataX} in the 
lowest bin. While we believe the low bin in $H_+(\qsq)$ represents a 
very small fraction of the cross section for the decay and is, hence,
particularly susceptible to unanticipated backgrounds, we cannot rule out
that this effect originates in the physics of the real decay.

We will refer to this representation of the data
as the \emph{un-convolved} analysis since the vertical error bars in Figure~\ref{ucdataX} do not
represent the uncertainty in the form factors averaged within each bin boundary. This is because
we have not corrected for the considerable smearing between the various $\qsq$ bins by using
the deconvolution technique~\cite{qsq} discussed in the next section.
To the extent that the form factors vary smoothly, the un-convolved representation should still be faithful to the 
underlying form factors as was the case of the Monte Carlo study shown in 
Figure~\ref{ucpop}.  We include the un-convolved results since they show more \qsq{} bins than possible in our 
deconvolution analysis discussed in the next section.

\begin{table}[htp]
\caption{Summary of un-convolved results. This is a tabular summary of the data of Fig. \ref{ucdataX} multiplied by 1000.}
\begin{center}
\begin{tabular}{c|c|c|c|c|c}
 \qsq{} & $\qsq{}_{rms}$ & $H_+^2$ & $H_-^2$ & $H_0^2$  & $h_0 \times H_0$  \\ 
\hline \hline
0.169& 0.125 & 0.42 $\pm$ 0.09 & 2.02 $\pm$ 0.87 & 8.91 $\pm$ 1.71 & 4.36 $\pm$ 1.53 \\ 
0.272& 0.136 & 0.31 $\pm$ 0.07 & 1.75 $\pm$ 0.22 & 4.09 $\pm$ 0.54 & 2.53 $\pm$ 0.78 \\ 
0.400& 0.154 & 0.32 $\pm$ 0.05 & 1.62 $\pm$ 0.14 & 2.44 $\pm$ 0.32 & 1.03 $\pm$ 0.57 \\ 
0.536& 0.175 & 0.34 $\pm$ 0.05 & 1.72 $\pm$ 0.12 & 2.40 $\pm$ 0.24 & 1.93 $\pm$ 0.46 \\ 
0.670 & 0.196 & 0.49 $\pm$ 0.06 & 1.68 $\pm$ 0.12 & 1.54 $\pm$ 0.24 & 1.01 $\pm$ 0.42 \\ 
0.789 & 0.218 & 0.48 $\pm$ 0.12 & 1.24 $\pm$ 0.20 & 1.64 $\pm$ 0.40 & 0.92 $\pm$ 0.56 \\
\\ 
\end{tabular}
\end{center}
\label{unconvsum}
\end{table}

\mysection{Deconvolution Weighting}
Because of $\qsq$ smearing, the sum of the $\vec P_+$ weights in a given reconstructed $\qsq$ bin will depend
on both the underlying $H^2_+({}^1\qsq)$ for that bin as well as 
the underlying $H^2_+({}^{i \ne 1}\qsq)$ form factor for all other bins.\footnote{
Throughout this discussion we will use superscripts for reconstructed $\qsq$ bin numbers and subscripts
for true $\qsq$ bin numbers.} Hence the old Eq.~\ref{secondweight} correction becomes the 
non-local version given by Eq.~\ref{nonlocal} where for notational simplicity we just consider two $\qsq$ bins. 

\begin{equation}
f_+({}^1q^2) = {}^1\vec P_+ \cdot {}^1\vec D  
             = {}^1w_1 \frac{H_+^2(q_1^2)}{(\tilde H_1^+)^2}
             + {}^1w_2 \frac{H_+^2(q_2^2)}{(\tilde H_2^+)^2} \,,
\label{nonlocal}
\end{equation}
where ${}^1w_1 = {}^1\vec P_+ \cdot {}^1\vec M_1$ is the sum of the Monte Carlo weights
that reconstruct in $\qsq$ bin 1 when generated in $\qsq$ bin 1 and 
${}^1w_2 = {}^1\vec P_+ \cdot {}^1\vec M_2$ is the sum of the Monte Carlo weights
that reconstruct in $\qsq$ bin 1 when generated in $\qsq$ bin 2.
Eq.~\ref{nonlocal} can be generalized to Eq.~\ref{matrixform}.

\begin{equation}
\left( {\begin{array}{*{20}c}
   {f_+({}^1q^2)}  \\
   {f_+({}^2q^2)}  \\
\end{array}} \right) = \left( {\begin{array}{*{20}c}
   {{}^1\vec P_+ \cdot {}^1\vec D}  \\
   {{}^2\vec P_+ \cdot {}^2\vec D}  \\
\end{array}} \right) = \left( {\begin{array}{*{20}c}
   {\frac{{}^1w_1}{(\tilde H_1^+)^2}} & {\frac{{}^1w_2}{(\tilde H_2^+)^2}}  \\
   {\frac{{}^2w_1}{(\tilde H_1^+)^2}} & {\frac{{}^2w_2}{(\tilde H_2^+)^2}}  \\
\end{array}} \right)\left( {\begin{array}{*{20}c}
   {H_+^2(q_1^2)}  \\
   {H_+^2(q_2^2)}  \\
\end{array}} \right) \,.
\label{matrixform}
\end{equation}
The solution to Eq.~\ref{matrixform} is given by Eq.~\ref{matrixsolution}.

\begin{equation}
\left( {\begin{array}{*{20}c}
   H_+^2(q_1^2)  \\
   H_+^2(q_2^2)  \\
\end{array}} \right) = \left( {\begin{array}{*{20}c}
   {\frac{{}^1w_1}{(\tilde H_1^+)^2}} & {\frac{{}^1w_2}{(\tilde H_2^+)^2}}  \\
   {\frac{{}^2w_1}{(\tilde H_1^+)^2}} & {\frac{{}^2w_2}{(\tilde H_2^+)^2}}  \\
\end{array}} \right)^{- 1} \left( {\begin{array}{*{20}c}
   {{}^1\vec P_+ \cdot {}^1\vec D}  \\
   {{}^2\vec P_+ \cdot {}^2\vec D}  \\
\end{array}} \right) \,.
\label{matrixsolution}
\end{equation}
This solution is equivalent to weighting the data using the weights given by Eq.~\ref{deconweights}.

\begin{equation}
\left( {\begin{array}{*{20}c}
   {\vec \rho_1^+} \\
   {\vec \rho_2^+} \\
\end{array}} \right) = \left( {\begin{array}{*{20}c}
   {\frac{{}^1w_1}{(\tilde H_1^+)^2}} & {\frac{{}^1w_2}{(\tilde H_2^+)^2}}  \\
   {\frac{{}^2w_1}{(\tilde H_1^+)^2}} & {\frac{{}^2w_2}{(\tilde H_2^+)^2}}  \\
\end{array}} \right)^{- 1} \left( {\begin{array}{*{20}c}
   {{}^1\vec P_+} \\
   {{}^2\vec P_+} \\
\end{array}} \right) \,.
\label{deconweights}
\end{equation}

In order to obtain $H_+^2(q_1^2)$, for example,  
one weights each event by $\vec \rho_1^{+}$. By Eq.~\ref{deconweights}, $\vec \rho_1^{+}$
is constructed from a sum of the vector ${}^1\vec P_+$ that contains the
angular weights for events that reconstruct in $\qsq$ bin 1
and the vector ${}^2\vec P_+$ that contains the
angular weights for events that reconstruct in $\qsq$ bin 2.  
 
\begin{figure}[htp]
\begin{center}
\includegraphics[height=3.in]{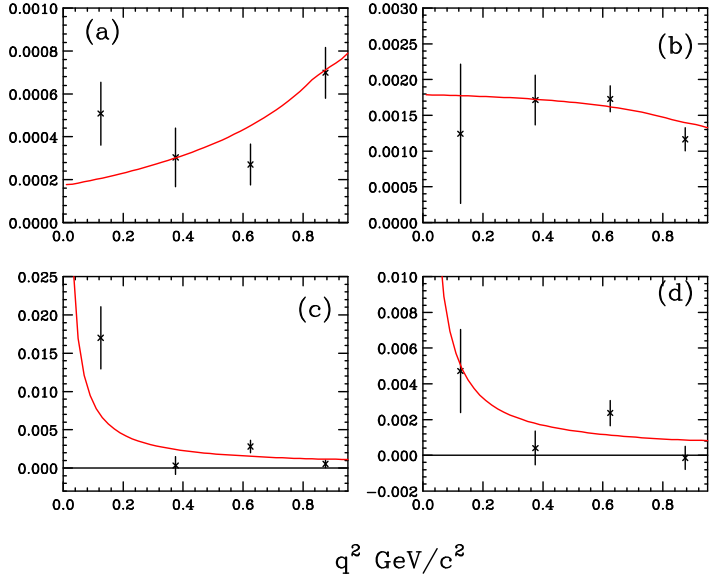}
\end{center}
\caption{
Non-parametric form factors obtained for deconvolved data. The plots are:
(a)~$H_+^2(\qsq)$,
(b)~$H_-^2(\qsq)$,
(c)~$H_0^2(\qsq)$, and
(d)~$h_0(\qsq) \times H_0(\qsq)$.
\label{decon}}
\end{figure}
\vskip .5in
\begin{table}[htp]
\caption{Summary of deconvolved results.  This is a tabular summary of the data of Fig. \ref{decon} multiplied by 1000.}
\begin{center}
\begin{tabular}{c|c|c|c|c}
 \qsq{} & $H_+^2$ & $H_-^2$ & $H_0^2$  & $h_0 \times H_0$  \\
\hline \hline 
0.125 & 0.508 $\pm$ 0.15 & 1.24 $\pm$ 0.97 & 17.00 $\pm$ 4.05 & 4.72 $\pm$ 2.33 \\ 
0.375 & 0.304 $\pm$ 0.14 & 1.71 $\pm$ 0.35 & 0.33 $\pm$ 1.17 & 0.41 $\pm$ 0.95 \\ 
0.625 & 0.270 $\pm$ 0.10 & 1.73 $\pm$ 0.18 & 2.82 $\pm$ 0.82 & 2.37 $\pm$ 0.71 \\ 
0.875 & 0.698 $\pm$ 0.12 & 1.16 $\pm$ 0.16 & 0.54 $\pm$ 0.49 & -0.14 $\pm$ 0.65 \\
\end{tabular}
\end{center}
\label{deconvsum}
\end{table}

Figure~\ref{decon} and Table~\ref{deconvsum} show the result of a four $\qsq$ bin deconvolution of the data. Only
four bins are used since as the $\qsq$ smearing exceeds the bin separation
the ``resolution" matrix that is inverted in Eq.~\ref{deconweights} becomes increasingly
more singular resulting in greatly inflated error bars as well as 
strong negative correlations appearing between adjacent bins~\cite{qsq}.
Essentially the same features that appear in the un-convolved representation of the data
seen in Figure~\ref{ucdataX} appear in Figure~\ref{decon} but on a coarser scale. There
is a small overshoot in the first bin of both $H^2_+$ and $H^2_0$, but generally both 
the shape and relative normalization of the form factors are a reasonable match to the model of
Reference~\cite{formfactor}.

We have estimated systematic errors by studying the stability of the results to changes in 
the number of angular bins, changes in the analysis cuts, and changes in our assumed
background level. For the three form factor products describing the \krzmndk{} component 
-- $H_+^2(\qsq)$, $H_-^2(\qsq)$, and $H_0^2(\qsq)$ -- the systematic errors are estimated to be
less than 20\% of the statistical error apart from the first \qsq{} bin  for $H^2_+ (\qsq{})$, where
we assess a systematic error equal to the statistical error. This first \qsq{} bin also requires a large ($\approx 30\%$)
background subtraction. For the $h_0(\qsq) \times H_0(\qsq)$ form factor product describing the s-wave interference
we assess a systematic error equal to 30\% of the statistic error.  The background subtraction for the 
$h_0(\qsq) \times H_0(\qsq)$ form factor product is also large ($\approx 50 \%$) for the first three $\qsq{}$ bins.
The errors quoted in Tables \ref{unconvsum} and \ref{deconvsum} do not include these systematic errors.

\mysection{Summary}

We presented the first non-parametric analysis of the helicity basis form
factors that control the decay $\kpimndk$. We used a projective weighting technique that allows one to 
determine the helicity form factor products by weighted histograms rather than
likelihood
based fits. This method should prove to be a
valuable technique for the $e^+e^-$ charm factories that have much better \qsq{}
resolution. The non-parametric technique can also be used for other
four body semileptonic decays like $D^+ \rightarrow \rho\; \ell^+ \nu_\ell$
and \philndk. 
We presented both an un-convolved and deconvolved representation of the data. Both
representations were a reasonable match to the spectroscopic pole dominance assumption with form
factor ratios given in~\cite{formfactor}.  The $\qsq$ dependence of the 
form factor governing the $s$-wave contribution to $\kpimndk$ is also studied for the first time. We
find that the shape of the $s$-wave form factor ($h_0(\qsq)$) is reasonably consistent with the $H_0(\qsq)$ form factor as
assumed in Reference~\cite{formfactor}.

\mysection{Acknowledgments}
We wish to acknowledge the assistance of the staffs of Fermi National
Accelerator Laboratory, the INFN of Italy, and the physics departments
of the collaborating institutions. This research was supported in part
by the U.~S.  National Science Foundation, the U.~S. Department of
Energy, the Italian Istituto Nazionale di Fisica Nucleare and
Ministero dell'Universit\`a e della Ricerca Scientifica e Tecnologica,
the Brazilian Conselho Nacional de Desenvolvimento Cient\'{\i}fico e
Tecnol\'ogico, CONACyT-M\'exico, the Korean Ministry of Education, and
the Korean Science and Engineering Foundation.

\end{document}